\documentclass{ws-ijmpa-v2}
\usepackage[super,compress]{cite}
\usepackage{hyperref}
\hypersetup{colorlinks,urlcolor=black,citecolor=black,linkcolor=black,filecolor=black}
\usepackage{breakurl}
\usepackage{graphicx,units}
\begin{document}
\markboth{Gabriele Honecker}{From Type II string theory towards BSM/dark sector physics}

%
\catchline{}{}{}{}{}
%

\title{From Type II string theory towards BSM/dark sector physics}

\author{Gabriele Honecker}

\address{Institut f\"ur Physik (WA THEP), Johannes Gutenberg-Universit\"at Mainz\\
D - 55099 Mainz, Germany\\
Gabriele.Honecker@uni-mainz.de}

%

\maketitle


\begin{abstract}
Four-dimensional compactifications of string theory provide a controlled set of possible gauge representations accounting for 
BSM particles and dark sector components. In this review, constraints from perturbative Type II string compactifications 
in the geometric regime are discussed in detail and then compared to results from heterotic string compactifications and 
non-perturbative/non-geometric corners. As a prominent example, an open string realization of the QCD axion is presented.
The status of deriving the associated low-energy effective action in four dimensions is discussed and open avenues 
of major phenomenological importance are highlighted. As examples, a mechanism of closed string moduli stabilization by D-brane backreaction
as well as one-loop threshold corrections to the gauge couplings and balancing a low string scale $M_{\text{string}}$
with unisotropic compact dimensions are discussed together with implications on potential future new physics observations. 
For illustrative purposes, an explicit example of a globally consistent D6-brane model with MSSM-like spectrum on
$T^6/(\mathbb{Z}_2 \times \mathbb{Z}_6 \times \Omega {\cal R})$ is presented.

\end{abstract}

\ccode{PACS numbers: 11.25.Wx, 11.25.Hf, 11.30.Fs, 12.60.-i, 14.70.Pw, 14.80.Va}


\tableofcontents

\section{Introduction}\label{S:Intro}	

Even though string theory remains  to date the arguably most successful framework for a unified description of Quantum Field Theory (QFT) and General Relativity - 
despite systematic extensive computer-aided searches such as in Refs.~\citen{Dijkstra:2004ym,Honecker:2004kb,Dijkstra:2004cc,Gmeiner:2006vb,Gmeiner:2007we,Gmeiner:2007zz,Gmeiner:2008xq,Robinson:2008qv,Rizos:2010sd,Rizos:2011ib,Anderson:2011ns,Anderson:2012yf,Honecker:2012qr,Anderson:2013xka,Nibbelink:2013lua,Ecker:2014hma,Nibbelink:2015vha,Ecker:2015vea} -  four dimensional string vacua with {\it exactly} the Standard Model or some GUT
particle content as well as gauge and Yukawa couplings remain elusive. While the searches for chiral particle physics spectra are limited by the knowledge of (mostly) topological data for suitable six-dimensional compact backgrounds, in particular Calabi-Yau threefolds, deriving the low-energy effective action further necessitates a more detailed
knowledge of the compact algebraic and differential geometry as well as techniques to quantize strings on such curved backgrounds.

One might thus argue that string theory is trapped between the juxtaposition of ``anything goes'' in terms of the assumed huge landscape of string vacua\cite{Vafa:2005ui,Schellekens:2015zua} and 
``nothing goes'' due to the lack of any explicitly known fully-fledged realistic string vacuum. As argued in this brief review, however, Beyond the Standard Model (BSM) and dark sector physics is considerably constrained within string theory compactifications as compared to purely field theoretical models, providing a 
guideline to search for new physics phenomena compatible with some UV completion. The focus here lies on representation theory and expanding the low-energy effective action
in explicitly known corners of the string landscape and is thus complementary to landscape arguments such as in Ref.~\citen{Dienes:2008rm}.

This review article is organized as follows: in section~\ref{S:Pert-II} generic model building rules from perturbative Type II superstring theories are presented 
with special emphasis on possibly allowed BSM/dark sector physics, which is compared in section~\ref{S:Het} with model building rules from heterotic string theories and non-perturbative regimes.
As two possible gateways to the dark sector, at first open string axions - in particular as models for the QCD axion - are discussed in section~\ref{S:Axions}, while one-loop gauge threshold corrections in Type II string theory and the relation to dark photons and Z' bosons for low values of $M_{\text{string}}$
 are presented in section~\ref{S:1-Loop}.
Section~\ref{S:Conclusions} contains the conclusions and outlook.

\section{Geometric Compactifications of Perturbative Type II Strings and D-Branes}\label{S:Pert-II}

In this section, we briefly review the state-of-the-art in Type IIA orientifold model building with D6-branes, paying special attention to its 
limitations on new BSM and dark sector states.
While the discussion focuses on the Type IIA case with D6-branes
for the sake of the geometric intuition and of the clarity of the argument, it has been conjectured that generalizing 
the T-duality arguments from tori and toroidal orbifolds of Refs.~\citen{Blumenhagen:2000ea,Angelantonj:2002ct} to mirror symmetry for Calabi-Yau threefolds leads to dual Type IIB orientifold models
with either magnetized D9/D5- or D7/D3-brane systems, see e.g. Refs.~\citen{Blumenhagen:2006ci,Ibanez:2012zz} for a comprehensive discussion and extended lists of references on perturbative Type IIB orientifold model building.

\subsection{Consistency conditions}\label{Ss:Consistency}

The string theoretic consistency conditions within the Type II string theory language can be grouped into two topological constraints and one using differential geometry:
\begin{enumerate}
\item
{\bf RR tadpole cancellation} ensures that the RR charges among D-branes and O-planes cancel along compact directions. 
Since the O-plane charges are fixed by the choice of an (antiholomophic) involution ${\cal R}$ along the six compact dimensions  accompanying (in Type IIA string theory)
the worldsheet parity operation $\Omega$, the cycles $\Pi$ on which D-branes can wrap together with their multiplicity or `stack size' $N$ are constrained.
In terms of D$6_a$-branes and O6-planes wrapping compact three-cycles this is described by:
\begin{equation}\label{Eq:RRtcc}
\sum_a N_a \left( \Pi_a + \Pi_a' \right) - 4 \, \Pi_{O6} =0,
\end{equation}
where $\Pi_a' \equiv {\cal R}(\Pi_a)$ is the orientifold image of the three-cycle $\Pi_a$.
\item
The additional $\mathbb{Z}_2$-valued information carried by D-branes beyond their homology class\cite{Witten:1998cd} contained in the RR tadpole cancellation conditions 
of Eq.~\eqref{Eq:RRtcc}
is captured by the {\bf K-theory constraints}, which are customarily formulated in terms of the absence of a field theoretical $SU(2) \simeq USp(2)$ anomaly\cite{Witten:1982fp,Uranga:2000xp,MarchesanoBuznego:2003axu},
\begin{equation}\label{Eq:Kconstr}
\Pi_{USp(2)} \circ \sum_a N_a \Pi_a = 0 \text{ mod } 2
\qquad \forall \text{  probe } USp(2),
\end{equation}
which is here for D6-branes written in terms of topological intersection numbers between three-cycles counting the number of chiral fermions in bifundamental representations.
To verify these constraints in a given model, one first has to rewrite the symplectic basis of three-cycles in terms of orientifold-even, $\Pi^+_i$,
 and orientifold-odd, $\Pi^-_i$, ones  ($i=0, \ldots, h^{21}$) and then classify for which three-cycles $\Pi^+_c=\sum_i X^i_c \Pi^+_i$ 
 (with wrapping numbers $X^i_a \in \mathbb{Z}$)
 D$6_c$-branes carry dimensionally enhanced gauge groups $USp(2N_c)$ or $SO(2N_c)$. 
\item
To ensure the stability of the string vacuum at the string scale $M_{\text{string}}$, {\bf supersymmetry} (SUSY) is imposed.\footnote{For $M_{\text{string}}$ as low as ${\cal O}$(TeV), one can argue that SUSY of the string vacuum is not mandatory but that e.g. Kaluza-Klein or winding modes should become visible in state-of-the-art collider experiments, e.g. at the LHC.~\cite{Antoniadis:1998ig,Shiu:1998pa,Accomando:1999sj,Cullen:2000ef,Kiritsis:2003mc,Burikham:2004su,Chemtob:2008cb,Anchordoqui:2009mm,Hashi:2011cz,Anchordoqui:2014wha,Anastasopoulos:2014lpa,Celis:2015eqs} }
 In terms of D6-branes and O6-planes wrapped around three-cycles, the SUSY condition is equivalent to all wrapped three-cycles being {\it special Lagrangian} ({\it sLag})  with the same calibration,
\begin{equation}\label{Eq:sLag}
J^{\text{K\"ahler}}_{1,1}|_{\Pi_a} =0,
\qquad
{\rm Re}(\Omega_3)|_{\Pi_a}>0,
\quad
{\rm Im}(\Omega_3)|_{\Pi_a}=0,
\end{equation}
 where $J^{\text{K\"ahler}}_{1,1}$ denotes the K\"ahler form and $\Omega_3$ the holomorphic volume form of 
 the compact Calabi-Yau threefold used for the compactification.
\end{enumerate}

In practice, the three types of stringy consistency conditions are straightforwardly computed for factorisable torus backgrounds, $T^6=(T^2)^3$, and orbifolds thereof, $(T^2)^3/\Gamma$
(with $\Gamma= \mathbb{Z}_N$ or $\mathbb{Z}_N \times \mathbb{Z}_M$ with(out) discrete torsion), on which 
a full classification of three-cycles with typical Betti number $b_3 \sim {\cal O}(10-50)$ is possible and for which Conformal Field Theory (CFT) methods can be invoked\cite{Berkooz:1996km,Blumenhagen:1999md,Blumenhagen:1999ev,Forste:2000hx} to not only cross-check the RR tadpole cancellation conditions in terms of vacuum amplitudes, but most importantly to determine the set of probe D-branes with $USp(2)$ gauge groups for the K-theory constraints.\cite{Uranga:2000xp,Forste:2010gw,Honecker:2011sm,Honecker:2013sww,Ecker:2014hma}. Generalizing these results to so-called non-factorizable tori, e.g. $T^6=(T^3)^2$ or $T^3 \times T^1 \times T^2$, or orbifolds thereof is possible whenever the {\it sLag} cycles can be rewritten in terms of a factorized geometry, as was recently noticed\cite{Seifert:2015fwr,Berasaluce-Gonzalez:2016kqb,Berasaluce:2016-2} when extending the first  CFT computations\cite{Blumenhagen:2004di,Forste:2007zb} to intersecting generic D6-branes with chiral spectra on non-factorizable $T^6/(\mathbb{Z}_4 \times \Omega {\cal R})$ backgrounds.
For generic Calabi-Yau threefolds as compact backgrounds, already determining the overall {\it sLag} three-cycle $\Pi_{O6}$ wrapped by the O6-planes in Eq.~\eqref{Eq:RRtcc}
is challenging\cite{Palti:2009bt} with - to our best knowledge - no known strategy for classifying probe D-branes with $USp(2)$ (and not $SO(2)$) gauge factors up to now.

\subsection{Massless spectrum}\label{Ss:Spectrum}

Generic three-cycles $\Pi_a$ wrapped by stacks of $N_a$ D$6_a$-branes provide $U(N_a)$ gauge factors, while orientifold-even 
three-cycles $\Pi_c^+$ support either $USp(2N_c)$ or $SO(2N_c)$ gauge groups of identical rank but enhanced dimension.
The {\it chiral} spectrum can be straightforwardly computed from topological intersection numbers between three-cycles as summarized in table~\ref{Tab:ChiralSpectrum},
\begin{table}[ph]
\tbl{Chiral spectrum of $\prod_x U(N_x) \times \prod_y USp/SO(2M_y)$.}
{\begin{tabular}{@{}cccc@{}} \toprule
Representation & Multiplicity \\ \colrule
$({\bf N}_a,\overline{\bf N}_b)$ &$ \Pi_a \circ \Pi_b$ \\
$({\bf N}_a,{\bf N}_b)$ &$ \Pi_a \circ \Pi_b'$ \\
$({\bf Anti}_a)$ & $\frac{\Pi_a \circ \Pi_a' + \Pi_a \circ \Pi_{O6} }{2}$\\
$({\bf Sym}_a)$ & $\frac{\Pi_a \circ \Pi_a' - \Pi_a \circ \Pi_{O6} }{2}$\\
$({\bf N}_a,{\bf 2M}_c)$ & $\Pi_a \circ \Pi_c$
\\ \botrule
\end{tabular} \label{Tab:ChiralSpectrum}}
\end{table}
while the vector-like (massless and massive) spectrum can (for D6-branes on three-cycles) to date only be accounted for by means of CFT computations on tori or toroidal orbifold backgrounds.
There exist basically two complementary methods for counting {\it all} massless matter states:
\begin{enumerate}
\item
Any open string state can be explicitly constructed, if the compact background is sufficiently simple and the string quantization condition is known, and Chan-Paton labels can be associated
to the different representations under the gauge groups, see e.g. Refs.~\citen{Blumenhagen:1999md,Blumenhagen:1999ev,Forste:2000hx} and~\citen{Gmeiner:2008xq,Forste:2010gw} for the corresponding CFT computations with {\it bulk} and {\it fractional} D6-branes, respectively. Via this method, the chirality of each open string state and its localization along the compact directions is determined. 
\item
The one-loop corrections to gauge couplings, including the full tower of massless and massive string excitations, can be computed by magnetically gauging string vacuum amplitudes and expanding in the gauging.~\cite{Abouelsaood:1986gd,Bachas:1996zt,Lust:2003ky} 
The full amount of massless string states and the corresponding representations (without chirality assignments) can then be read off from the contributions
of the different open string sectors to the beta function coefficients, which appear as prefactors of the $\frac{1}{\varepsilon} + \gamma_E - {\rm ln}2  \simeq \ln \frac{M_{\text{string}}^2}{\mu^2}$ terms in dimensional regularisation $\int d\ell \, \ell^{\varepsilon} [\ldots]$ of the corresponding vacuum amplitudes.\cite{Akerblom:2007np,Blumenhagen:2007ip,Gmeiner:2009fb,Honecker:2011sm} 
\end{enumerate}
The method of Chan-Paton labels is of vital importance when setting up the framework to compute Yukawa and higher $m$-point couplings, while 
the advantage of the one-loop gauge threshold computation is that (otherwise cumbersome) sign factors for distinguishing symmetric and antisymmetric
representations are easily kept track of, and 
that extended computer scans and classifications of massless spectra in global D-brane models can be performed (with suitably powerful computers or clusters), when 
combining with the signs of intersection numbers to determine net-chiralities.
 
The first important message to take away at this point in view of BSM and/or dark sector physics
is that any endpoint of an open string in perturbative Type II string theory transforms in the (anti)fundamental representation of 
the gauge group, which is supported on the (stack of) D-brane(s) the endpoint is confined to. Taking into account the orientifold projection, the following non-Abelian
representations (plus their conjugates) under any $U(N_a) \times U(N_b) \times USp(2M_c) \times SO(2M_d)$ group factor can appear:
\begin{eqnarray}
 & ({\bf N}_a,\overline{\bf N}_b),
 \quad
 ({\bf N}_a,{\bf N}_b),
 \quad
({\bf N}_{x \in \{a,b,\}},{\bf 2M}_{y \in \{c,d\}}),
\quad
({\bf 2M}_c,{\bf 2M}_d),
\nonumber\\
& ({\bf Adj}_{x \in \{a,b\}}),
\quad
({\bf Anti}_{x \in \{a,b,c,d\}}),
\quad
({\bf Sym}_{x \in \{a,b,c,d\}}),
 \end{eqnarray}
while higher dimensional representations, such as e.g. $({\bf Anti}_a,{\bf N}_b)$ or $({\bf N}_a,{\bf N}_b,{\bf N}_c)$, or spinorial representations, e.g. $({\bf 2^{M-1}}_{\pm})$, of $SO(2M)$ gauge factors can never occur. $SO(10)$ Grand Unified Theories (GUTs) or exceptional gauge groups are thus not accessible from geometrically engineered perturbative Type II string vacua, while Pati-Salam, left-right symmetric models and MSSM-like spectra arise naturally.

\subsubsection{$U(1)$ factors and discrete $\mathbb{Z}_n$ symmetries}\label{Sss:Zn}

The Abelian gauge factors within $\prod_x U(N_x)$ mix to form mass eigenstates, $U(1)_X = \sum_x q_x U(1)_x$ with 
coefficients $q_x \in \mathbb{Q}$. 
In four spacetime dimensions,  mass terms and couplings of ${\rm tr} F_x^{k,k\in\{1,2\}}$  to closed string two-forms ${\cal B}^{(i)}_2$ and their dual axions $\xi_i$
(with $d{\cal B}^{(i)}_2=m_i \ast_4 d\xi_i$ for $m_i \in \mathbb{Z}$ depending on the three-cycle basis for a given six-dimensional compact space),  which are in the Type IIA language complexifications of complex structure moduli and the dilaton, arise from the Chern-Simons action along the D6-branes,\cite{Ibanez:2001nd}
\begin{equation}\label{Eq:GS}
{\cal S}_{CS} \supset \sum_x \sum_{i=0}^{h^{21}}  \int_{\mathbb{R}^{1,3}} \left( 
Y^i_x \, {\cal B}^{(i)}_2 \wedge {\rm tr} F_x + 
X^i_x \, \xi_i {\rm tr} F_x \wedge F_x
\right)
,
\end{equation}
with three-cycle wrapping numbers $X^i_x, Y^i_x \in \mathbb{Z}$ stemming from the expansion $\Pi_x = \sum_{i=0}^3 [ X^i_x \Pi^+_i + Y^i_x \Pi^-_i]$ in orientifold-even and orientifold-odd parts introduced in section~\ref{Ss:Consistency}. 
A linear combination $U(1)_X$ remains massless and anomaly-free if the St\"uckelberg couplings to all two-forms in Eq.~\eqref{Eq:GS} vanishe, i.e. if
$\sum_x  q_x N_x Y^i_x=0$ for all $i$.

{\bf Classification of hypercharge embeddings:} the origin of all matter representations from pairs of open string endpoints constrains the possible massless linear combinations which can reproduce the charge assignments of the hypercharge of all left- and right-handed quarks and leptons. Up to exchange of orientifold image D-branes ($x \leftrightarrow x'$ for $x \in \{b,c,d\}$), 
there exist only four different possibilities\cite{Gmeiner:2005vz,Anastasopoulos:2012zu}:
\begin{equation}\label{Eq:possible-Ys}
(q_a,q_b,q_c,q_d) \in \left\{\left(\frac{1}{6},0,\frac{1}{2},\frac{1}{2} \right), \left(\frac{-1}{3},\frac{-1}{2},0,0 \right), 
\left(\frac{-1}{3},\frac{-1}{2},0,1 \right), \left(\frac{1}{6},\frac{1}{2},0,\frac{-3}{2} \right) \right\}
,
\end{equation}
to realize the Standard Model hypercharge and particle content on four (or three) stacks of D-branes with gauge groups $U(3)_a \times U(2)_b \times U(1)_c ( \times U(1)_d)$, or for 
the `standard embedding' $(\frac{1}{6},0,\frac{1}{2},\frac{1}{2})$ also $U(3)_a \times USp(2)_b \times U(1)_c (\times U(1)_d)$. 

In case of the `standard embedding' and non-rigid D-branes, the `right' symmetry $U(1)_c$ can arise from a breaking of a right-symmetric group $USp(2)_c$ or $SO(2)_c$ along some flat direction in the Wilson line and displacement moduli space. Since in this case $U(1)_c$ remains massless, these models also possess a gauged baryon-lepton number $(B-L)$ symmetry with $(q_a,q_b,q_c,q_d)
=(\frac{1}{3},0,0,1)$.\cite{Honecker:2004kb,Gmeiner:2008xq,Gmeiner:2009fb}

{\bf $\mathbb{Z}_n$ symmetries:}
The original $U(1)_x \subset U(N_x)$ factors, or equivalently the massive linear combinations thereof, remain as perturbative global symmetries in the low-energy effective action.
As expected in any quantum gravity framework,\cite{Kallosh:1995hi,Banks:2010zn} these continuous global symmetries are broken to discrete subgroups $\mathbb{Z}_n$ non-perturbatively, e.g. by D-brane instanton effects, provided that 
\begin{equation}
\sum_x k_x  N_x Y^i_x=0
\text{ mod }
n
\qquad 
\forall \quad i=0 \ldots h^{21}
,
\end{equation}
with integers $k_x$ such that ${\rm gcd}(n,k_a,k_b, \ldots)=1$.\cite{BerasaluceGonzalez:2011wy,Ibanez:2012wg,Anastasopoulos:2012zu,Honecker:2013sww} 
These discrete symmetries provide the ultimate selection rules on couplings in the low-energy effective field theory.
While the majority of examples to date consists of generation-independent $\mathbb{Z}_n$ symmetries, in Refs.~\citen{Honecker:2013sww,Honecker:2015ela}
a generation-dependent $\mathbb{Z}_2$ was found, which can account for highly suppressed off-diagonal Yukawa couplings. 

{\bf Peccei-Quinn symmetries:}
In 1977, a spontenously broken global $U(1)_{PQ}$ symmetry was proposed to solve the strong CP problem.\cite{Peccei:1977hh}
Within string theory models, it is natural to identify this $U(1)_{PQ}$ symmetry as one of the massive linear combinations (with mass $m_{U(1)_{PQ}} \sim M_{\text{string}}$) 
that are present as global symmetries at the perturbative level. Demanding that the QCD axion as well as the Higgs(es) and either right- or left-handed Standard Model particles are charged, 
boils down to two different possibilities for the `standard embedding' of the hypercharge, namely $U(1)_{PQ} = U(1)_b \subset U(2)_b$ or $U(1)_{PQ} = U(1)_c - U(1)_d$
as discussed in more detail in section~\ref{S:Axions}.
While non-perturbative effects will break such a $U(1)_{PQ}$ symmetry, some discrete subgroup can be preserved such as a $\mathbb{Z}_3$ gauge symmetry in the D6-brane model with MSSM spectrum\cite{Ecker:2015vea,Honecker:2015qba} displayed in section~\ref{Sss:MSSM-Spectrum}, see e.g. also Ref.~\citen{Abel:1995wk} for a field theoretic model with $\mathbb{Z}_3$ symmetry or Ref.~\citen{Cvetic:2015moa} in the context of F-theory.

\subsubsection{Example: MSSM on $T^6/(\mathbb{Z}_2 \times \mathbb{Z}_6 \times \Omega{\cal R})$}\label{Sss:MSSM-Spectrum}

To illustrate the above features, we present here an example of a globally consistent D6-brane model on the $T^6/(\mathbb{Z}_2 \times \mathbb{Z}_6
\times \Omega {\cal R})$ orientifold with discrete torsion\cite{Ecker:2015vea,Honecker:2015qba},
which has Hodge numbers $(h^{11},h^{21})=(3_{\text{bulk}} + 8_{\mathbb{Z}_6'} + 8_{\mathbb{Z}_3} \, , \, 1_{\text{bulk}}+2_{\mathbb{Z}_6} + 2_{\mathbb{Z}_3} + (6+4+4)_{\mathbb{Z}_2})$ for the factorizable background lattice $(A_1)^2 \times A_2 \times A_2$. 
The {\it a priori} six different choices of orientations of the $(T^2)^3$ lattice under the antiholomorphic involution ${\cal R}$ can be reduced to two\cite{Ecker:2014hma}, i.e.  $T^2_{(1)}$ is either of rectangular or of tilted shape, and due to the discrete torsion phase one O6-plane orbit $\Omega{\cal R}(\mathbb{Z}_2^{(i)})$ has to be of exotic type\cite{Blumenhagen:2005tn,Forste:2010gw}, 
with two inequivalent choices, $\Omega{\cal R}$ or $\Omega{\cal R}\mathbb{Z}_2^{(3)}$, consistent with SUSY D6-branes on the so-called {\bf aAA}-lattice orientation.

The $2 \times (h^{21} + 1)_{\text{bulk} + \mathbb{Z}_2}=30$ three-cycles from the bulk and $\mathbb{Z}_2$-twisted sectors can be expanded in a symplectic basis of 15 orientifold-even and 15 orientifold-odd three-cycles as detailed in Ref.~\citen{Ecker:2015vea}. For the choice of $\Omega{\cal R}\mathbb{Z}_2^{(3)}$ as the exotic O6-plane orbit, SUSY solutions to the RR tadpole cancellation condition in Eq.~\eqref{Eq:RRtcc} have the unusual feature that all D6-branes are parallel to the $\Omega{\cal R}(\mathbb{Z}_2^{(1)})$-invariant planes
along $T^2_{(1)}$ and at non-trivial angles along $T^4_{(1)} \equiv T^2_{(2)} \times T^2_{(3)}$. Under the orientifold projection, $h^{11}$ splits 
into $(h^{11}_{\mathbb{Z}_6'})^+=4$ and $(h^{11}_{\text{bulk} + \mathbb{Z}_6' + \mathbb{Z}_3 })^-= 15$, which count multiplets containing closed string vectors and K\"ahler moduli, respectively.
The implications of all D6-branes wrapping the same one-cycle along $T^2_{(1)}$ 
on a potentially rather low value of the string scale $M_{\text{string}}$ in dependence of the radii $R_1^{(1)}, R_2^{(1)}$
 for this specific kind of configuration will be further discussed in section~\ref{S:1-Loop}.

The SUSY D6-brane configuration for a specific MSSM-like particle spectrum,\cite{Ecker:2015vea} 
which satisfies all RR tadpole cancellation conditions of Eq.~\eqref{Eq:RRtcc} and K-theory constraints of Eq.~\eqref{Eq:Kconstr} is displayed in table~\ref{Tab:D6-Config-MSSM}.
\begin{table}[h!]
\tbl{D6-brane configuration for a MSSM on $T^6/(\mathbb{Z}_2 \times \mathbb{Z}_6\times \Omega{\cal R})$.}
{\begin{tabular}{@{}ccccccc@{}} \toprule
Brane & $(n_i,m_i)$ & $\frac{\text{angle}}{\pi}$ & $\mathbb{Z}_2^{(i)}$ & $(\vec{\tau})$ & $(\vec{\sigma})$ & group
\\ \colrule
$a$ & (1,0;1,0;1,0) & (0,0,0) & $(+++)$ & (0,1,1) & (0,1,1) & $U(3)_a$  \\
$b$ & (1,0;-1,2;1,-2) & $(0,\frac{1}{2},\frac{-1}{2})$ & $(--+)$ & (0,1,0) & (0,1,0) & $USp(2)_b$ \\
$c$ &  (1,0;-1,2;1,-2) & $(0,\frac{1}{2},\frac{-1}{2})$ & $(-+-)$ & (0,1,1) & (0,1,1) & $U(1)_c$ \\
$d$ &  (1,0;-1,2;1,-2) & $(0,\frac{1}{2},\frac{-1}{2})$ & $(+--)$ & (0,0,1) & (0,0,1) & $U(1)_d$\\
$h$ &  (1,0;1,0;1,0) & (0,0,0) & $(--+)$ & (0,1,1) & (0,1,1) & $U(4)_h$
\\ \botrule
\end{tabular} \label{Tab:D6-Config-MSSM}}
\end{table}
The `standard embedding' of the hypercharge according to Eq.~\eqref{Eq:possible-Ys} constitutes the only massless Abelian gauge symmetry besides 
a $\mathbb{Z}_3$ symmetry that forms a remnant of the massive $U(1)_{PQ} \simeq U(1)_c-U(1)_d$ symmetry in this example.
The massless open string spectrum charged under the low-energy gauge group $\left(SU(3)_a \times SU(2)_b \times SU(4)_h \right)_{\times U(1)_Y}^{\times \mathbb{Z}_3}$ 
is displayed in table~\ref{Tab:Spectrum-MSSM}.
\begin{table}[h!]
\tbl{Massless spectrum of a MSSM example on $T^6/(\mathbb{Z}_2 \times \mathbb{Z}_6 \times \Omega{\cal R})$.}
{\begin{tabular}{@{}ccccccccccccc@{}} \toprule
sector & $ab$ & $ac$ & $ad$ & $ad'$ & $bc$ 
\\
matter & $3 \, ({\bf 3},{\bf 2};{\bf 1})_{\nicefrac{1}{6}}^{0}$ & $ 6 \, (\bar{\bf 3},{\bf 1};{\bf 1})_{\nicefrac{1}{3}}^{1}$
& $ 3\, ({\bf 3},{\bf 1};{\bf 1})_{\nicefrac{-1}{3}}^{1}$ & $ 3\, (\bar{\bf 3},{\bf 1};{\bf 1})_{\nicefrac{-2}{3}}^1$
& $\begin{array}{c} 3 \, ({\bf 1},{\bf 2};{\bf 1})_{\nicefrac{1}{2}}^1 \\ + 3 \, [({\bf 1},{\bf 2};{\bf 1})_{\nicefrac{1}{2}}^1 + h.c.]\end{array}$
\\ \colrule
sector & $ bd$ & $cd$ & $cd'$ 
\\
matter & $\begin{array}{c} 6 \, ({\bf 1},{\bf 2};{\bf 1})_{\nicefrac{-1}{2}}^1 \\ + 2 \, [({\bf 1},{\bf 2};{\bf 1})_{\nicefrac{-1}{2}}^1 + h.c.]\end{array}$
&  $\begin{array}{c} 3 \, ({\bf 1},{\bf 1};{\bf 1})_{0}^{1} \\ + 3 \, [({\bf 1},{\bf 1};{\bf 1})_{0}^{1} + h.c.]\end{array}$
&  $\begin{array}{c} 3 \, ({\bf 1},{\bf 1};{\bf 1})_{1}^{0} \\ + 3 \, [({\bf 1},{\bf 1};{\bf 1})_{1}^{0} + h.c.]\end{array}$
\\ \colrule
sector & $aa'$ & $bb+cc+dd$ & $dd'$ & $ah$ & $ah'$ 
\\
matter 
& $2 \, [(\bar{\bf 3}_A,{\bf 1};{\bf 1})_{\nicefrac{1}{3}}^0 + h.c.]$
& $ (5 + 4 + 5) \, ({\bf 1},{\bf 1};{\bf 1})_0^0$
& $[({\bf 1},{\bf 1};{\bf 1})_{1}^1 + h.c.]$
& $2 \, [({\bf 3},{\bf 1};\bar{\bf 4})_{\nicefrac{1}{6}}^0 + h.c.]$
& $[({\bf 3},{\bf 1};{\bf 4})_{1/6}^0 + h.c.]$
\\ \colrule
sector & $bh$ & $ ch'$ & $dh$ & $dh'$ & $hh'$
\\
matter & $ 3 \, ({\bf 1},{\bf 2};{\bf 4})_0^0$
& $ 6 \, ({\bf 1},{\bf 1};\bar{\bf 4})_{\nicefrac{-1}{2}}^2$
& $ 3\, ({\bf 1},{\bf 1};\bar{\bf 4})_{1/2}^2$
& $ 3\, ({\bf 1},{\bf 1};{\bf 4})_{\nicefrac{1}{2}}^2$
& $ 2 \, [({\bf 1},{\bf 1};{\bf 6}_A)_0^0 + h.c.]$
\\ \botrule
\end{tabular} \label{Tab:Spectrum-MSSM}}
\end{table}
Since this model does not possess any gauged $(B-L)$ symmetry, the chiral multiplets in the $bc+bd$ sectors containing 
Higgs scalars and left-handed leptons can only be distinguished by their superpotential couplings to quarks and  leptons
as briefly discussed in section~\ref{Sss:BLO}. Similarly, the $cd$ sector consists of chiral multiplets which contain not only three generations of right-handed Weyl neutrinos, but also candidates for the QCD axion as discussed further in section~\ref{S:Axions}.

It is noteworthy that, even though the toroidal orbifold background contains singularities, the geometric engineering method provides exactly the same 
spectrum as expected in the smooth Calabi-Yau case, after neutral closed string blow-up and deformation modes have been used. This holds true even though some of the deformation moduli will be stabilized at the singular orbifold point via couplings to D6-branes as discussed further in section~\ref{Sss:ModuliStab}.
This observation on the matter spectrum is in contrast to heterotic orbifolds and Gepner models discussed in section~\ref{Ss:HetOrbifolds} and~\ref{Ss:Gepner}, respectively, where slight enhancements of gauge groups and representations as well as shifts of $U(1)$ charges in twisted sector can occur.
In the case of heterotic orbifold models, geometric moduli can moreover not be uniquely distinguished from matter fields since all originate on equal footing
 from closed strings.

\subsection{Towards the four-dimensional effective action}\label{Ss:Action}

While the chiral spectrum on intersecting D6-branes can be computed solely from topological intersection numbers of the corresponding three-cycles
on any Calabi-Yau threefold, already the massless vector-like matter spectrum requires a more refined knowledge of the compact six-dimensional background. The situation becomes even more challenging when trying to reproduce the orders of magnitude of and hierarchies among Standard Model or some GUT
gauge and Yukawa couplings.

The present techniques rely on combining dimensional reductions of the ten-dimensional supergravity (SUGRA) as well as Dirac-Born-Infeld (DBI)
and Chern-Simons (CS) actions along D-branes with CFT computations of scattering amplitudes for sufficiently simple toroidal (orbifold) backgrounds.
The scope and state-of-the-art of these techniques will be discussed individually, and in section~\ref{Sss:ModuliStab} 
 complex structure moduli stabilization at the orbifold point will be discussed.

\subsubsection{Dimensional reduction of SUGRA, DBI and CS actions}\label{Sss:SUGRA}

For any perturbative string theory, part of the tree-level effective action in four dimensions can be derived by dimensionally reducing the ten-dimensional 
SUGRA action. Expanding all massless closed string fields in terms of (cohomology classes of) differential forms along the compact space, or equivalently integrating the ten-dimensional fields over the dual cycles (or homology classes), 
yields the four-dimensional K\"ahler and complex structure moduli encoded in the K\"ahler form $J^{\text{K\"ahler}}_{1,1}$ and holomorphic volume form $\Omega_3$, closed string vectors as well as the four-dimensional metric and the dilaton, along with their axionic and fermionic partner fields as worked out in detail in Ref.~\citen{Grimm:2004ua}. 
Besides contributions to the moduli potential, the reduction of the Einstein-Hilbert term, 
\begin{equation}
{\cal S}_{\text{IIA}} \supset \frac{1}{2 \kappa_{10}}
 \int_{10{\rm D}} {\rm d}^{10}x \, \sqrt{-g_{10}} \, e^{-2 \phi_{10}} \, {\cal R}_{10}
 \quad \to \quad  
 \frac{1}{2 \kappa_{4}}
 \int_{\mathbb{R}^{1,3}} {\rm d}^4x \, \sqrt{-g_4} \, {\cal R}_4,
\end{equation}
provides an important relation among the characteristics of the string compactification and the four-dimensional Planck scale,
\begin{equation}\label{Eq:M-Pl-M-str}
\frac{M_{\text{Planck}}^2}{M_{\text{string}}^2} = \frac{4\pi}{g_{\text{string}}^2} \frac{{\rm Vol}(CY_3)}{\ell_s^6}
.
\end{equation}
The string length is here defined in terms of the Regge slope $\alpha'$
as $\ell_s \equiv 2 \pi \sqrt{\alpha'}$, the string coupling as \mbox{$g_{\text{string}} \equiv e^{\langle \phi_{10} \rangle}$}
and the ten- and four-dimensional gravitational coupling constants as \mbox{$\kappa_{10}^2 \equiv \frac{\ell_s^8}{4\pi}$}
and \mbox{$\kappa_4^2 \equiv M_{\text{Planck}}^{-2}$}, respectively.

In orientifolded Type II string theories, another part of the tree-level effective action in four dimensions can be obtained by dimensionally reducing 
the DBI and CS actions, see e.g. Refs.~\citen{Grimm:2011dx,Kerstan:2011dy} in the context of  D6-branes,
\begin{equation}\label{Eq:DBI}
{\cal S}_{\text{D}6_a} \supset 
- \frac{1}{8\pi \ell_s^3} \int_{7{\rm D}}{\rm d}^{7}x \, e^{-\phi_{10}} \, F^a_{MN}F^{MN}_a
\quad \to \quad 
- \frac{1}{4 \, g_{a,\text{tree}}^2} \int_{\mathbb{R}^{1,3}} {\rm d}^4x \, F^a_{\mu\nu} F^{\mu\nu}_a
.
\end{equation}
Generalizing this term from a single D$6_a$-brane carrying a $U(1)_a$ gauge group to non-Abelian gauge groups gives the relation
\begin{equation}\label{Eq:g-tree}
\frac{4\pi}{g_{a,\text{tree}}^2} = \frac{1}{4  c_a  k_a \,  g_{\text{string}}} \, \frac{{\rm Vol}(\Pi_a)}{\ell_s^3}
 \text{ with } 
c_a=\left\{\begin{array}{cc} 1 & \text{bulk} \\ 2 & \text{fractional} \\ 4 & \text{rigid} \end{array}\right.
,
k_a= \left\{\begin{array}{cc} 1 & SU(N_a) \\ 2  & USp/SO(2N_a) \end{array}\right.
,
\end{equation}
where the factor $c_a$ accounts for the fact that on toroidal orbifolds, the unimodular basis of three-cycles 
has the form $\Pi_a = (\Pi_a^{\text{bulk}}+ \Pi_a^{\mathbb{Z}_2})/c_a$ with $\Pi_a^{\mathbb{Z}_2}$ a collection 
of exceptional three-cycles, which have zero volume at the singular orbifold point.

While the dimensional reduction is the easiest accessible technique using algebraic geometry on generic Calabi-Yau spaces,
it is limited to tree-level effects stemming from massless string modes only. Moreover, the generalization of the full DBI action 
to stacks of D-branes and thus non-Abelian gauge groups is not known, and interactions originating from intersections of 
different stacks of D-branes, in particular Yukawa couplings, cannot be computed by this method. 

For heterotic string theories, besides the tree-level supergravity action, the one-loop Green-Schwarz counter terms\cite{Green:1984sg} required for
the cancellation of hexagonal gravitational, gauge and mixed anomalies in ten dimensions are known. These are matched by S-dual tree-level terms in the DBI and CS action within Type II string compactifications with D-branes.\cite{Blumenhagen:2005pm,Blumenhagen:2005zh,Gmeiner:2005vz}
A brief comparison with heterotic model building, including also F-theory, is provided below in section~\ref{S:Het}.
%

\subsubsection{CFT \& beyond leading order}\label{Sss:BLO}

Whenever the compactification background is simple enough and the string quantization on this space is explicitly known, 
CFT techniques can be employed to derive the one-loop corrections to the gauge couplings and the tree-level superpotential
involving matter fields. Both types of contributions to the effective action are protected by perturbative non-renormalization theorems 
for supersymmetric field theories, but can receive non-perturbative contributions, e.g. from D-brane instantons, which 
can at least in principle also be computed by means of CFT. Also the fields, in particular the geometric moduli and the dilaton, 
entering the low-energy effective field theory might be subject to field redefinitions beyond leading order.

{\bf One-loop corrections to the gauge couplings:}
the tree-level and one-loop contributions to the gauge couplings can be computed in the same way as higher $m$-point couplings
by inserting two vertex operators at the boundaries of open string worldsheets, which are disks at tree level and annuli and M\"obius strips at one-loop level.
The functional dependence on some closed string modulus is probed by adding the corresponding vertex operator in the bulk of the worldsheet. 
In Ref.~\citen{Lust:2004cx}, it was shown that (up to an undetermined numerical prefactor), the tree level result for the gauge coupling of a D6-brane
agrees with Eq.~\eqref{Eq:g-tree} derived from dimensional reduction of the DBI action. 

Fortunately, the one-loop corrections to the gauge couplings can also be computed in an alternative, somewhat simpler way without using vertex operators by instead magnetically gauging open string vacuum amplitudes\cite{Bachas:1996zt,Antoniadis:1999ge,Abouelsaood:1986gd,Callan:1986bc,Bachas:1992bh} with annulus and M\"obius strip topologies along $\mathbb{R}^{1,3}$ and expanding in the gauging,
\begin{equation}
b_a {\rm ln} \frac{M_{\text{string}}^2}{\mu^2} + \Delta_a = \sum_b \left[ {\cal T}^{\cal A}(\text{D}6_a,\text{D}6_b) + {\cal T}^{\cal A}(\text{D}6_a,\text{D}6_{b'}) 
\right] + {\cal T}^{\cal M}(\text{D}6_a,{\cal O}6)
,
\end{equation}
where the usual identification $\frac{1}{\varepsilon} + \gamma_E - {\rm ln}2 \simeq {\rm ln} \frac{M_{\text{string}}^2}{\mu^2}$ after dimensional regularization as anticipated in section~\ref{Ss:Spectrum} has been made.
The results for intersecting {\it bulk} D6-branes on factorizable toroidal backgrounds or orbifolds thereof where first derived in Ref.~\citen{Lust:2003ky,Akerblom:2007np} for vanishing vacuum expectation values of the continuous open string moduli. Partial results for {\it rigid} D6-branes were obtained in Ref.~\citen{Blumenhagen:2007ip}, and the formulas were systematically completed in Refs.~\citen{Gmeiner:2009fb,Honecker:2011sm} for {\it bulk}, {\it fractional} and 
{\it rigid} D6-branes.
For {\it rigid} D6-branes, the contributions are summarized in the last column of table~\ref{Tab:Betas+1-Loop} with $V_{ab}^{(i)}=\frac{R_1^{(i)}}{R_2^{(i)}} n^i_an^i_b + 
\frac{R_2^{(i)}}{R_1^{(i)}}(m^i_a+b_in^i_a)(m^i_b+b_in^i_b)$ for generic untilted ($b_i=0$) or untilted ($b_i=\frac{1}{2}$) tori 
or $V_{ab}^{(i)}=\frac{2n^i_an^i_b +n^i_am^i_b +m^i_an^i_b +2m^i_am^i_b  }{\sqrt{3}}$ for hexagonal tori, $I_{ab}$ denoting toroidal intersection numbers
and $I_{ab}^{\mathbb{Z}_2}$ $\mathbb{Z}_2$-invariant intersection numbers dressed with sign factors due to $\mathbb{Z}_2$ eigenvalues and discrete Wilson lines $\tau_i$ as detailed  in e.g. appendix~A of Ref.~\citen{Honecker:2012qr}. The 
lattice sums appearing in the gauge thresholds are defined\cite{Honecker:2011sm} as $\Lambda_{0,0}(v)=\ln(2\pi v V \eta^4(iv))$ and 
$\Lambda_{\tau,\sigma \neq 0,0} =\ln \bigl| e^{-\frac{\pi \sigma^2 v}{4}} \frac{\vartheta_1 (\frac{\tau-i\sigma v}{2},iv)}{\eta(iv)}\bigr|^2$,
with $\sigma^{ab}_i \equiv |\sigma^a_i - \sigma^b_i| \in \{0,1\}$ and $v_i$ the two-torus volumes in units of $\alpha'$.
\begin{table}[ph]
\tbl{Annulus contributions to beta function coefficients, K\"ahler metrics and 1-loop threshold corrections for rigid D6-branes.}
{\begin{tabular}{@{}ccccc@{}} \toprule
$\frac{\text{angle}}{\pi}$ & $b^{\cal A}_{SU(N_b)} (= \tilde{b}^{\cal A}_{SU(N_b)}\delta_{\sigma^{a}_i}^{\sigma^b_i}
\delta_{\tau^{a}_i}^{\tau^b_i}) $
& $K_{({\bf N}_a,\overline{\bf N}_b)}$ & $\Delta_{SU(N_a)}^{\cal A}$
\\ \colrule
$(0,0,0)$ & $- \frac{N_b \sum_{i=1}^3 I_{ab}^{\mathbb{Z}_2^{(i)},(j\cdot k)}}{4} $ 
& $e^{\phi_4} \sqrt{\frac{2\pi V_{ab}^{(i)}}{v_jv_k}} $
& $-\sum_{i=1}^3 \tilde{b}_{SU(N_a)}^{{\cal A},(i)} \Lambda_{\tau_i^{ab},\sigma_i^{ab}} (v_i)$
\\
$(0,\phi,-\phi)$ & $\frac{N_b \bigl(|I_{ab}^{(2 \cdot 3)}| -  I_{ab}^{\mathbb{Z}_2^{(1)},(2\cdot 3)}\bigr) }{4} \delta_{\sigma^{a}_1}^{\sigma^b_1}
\delta_{\tau^{a}_1}^{\tau^b_1}$
& $e^{\phi_4} \sqrt{\frac{2\pi V_{ab}^{(1)}}{v_2v_3}}$
&  $\left\{\begin{array}{c} -\tilde{b}_{SU(N_a)}^{\cal A} \Lambda_{\tau_1^{ab},\sigma_1^{ab}}(v_1)\\
+ \frac{N_b \bigl( I_{ab}^{\mathbb{Z}_2^{(2)}} - I_{ab}^{\mathbb{Z}_2^{(3)}}  \bigr)\bigl( {\rm sgn} (\phi) -2 \phi \bigr)\ln2}{4} 
\end{array}\right.
$
\\
$(\phi^{(1)},\phi^{(2)},\phi^{(3)})$ & $\frac{N_b\bigl(I_{ab} + {\rm sgn} ( I_{ab}) \sum_{i=1}^3 I_{ab}^{\mathbb{Z}_2^{(i)}}\bigr)}{8}  $
& $e^{\phi_4} \sqrt{\prod_{i}
\frac{1}{v_i} \left( \frac{\Gamma(|\phi^{(i)}|)}{\Gamma(1-|\phi^{(i)}|)}
\right)^{-\frac{{\rm sgn}(\phi^{(i)})}{{\rm sgn}(I_{ab})}} }$
& $\frac{N_b \sum_{i=1}^3 I_{ab}^{\mathbb{Z}_2^{(i)}} \bigl( {\rm sgn} (\phi^{(i)}) -2 \phi^{(i)} + {\rm sgn}(I_{ab}) \bigr)
 \ln2 }{4}$
\\ \botrule
\end{tabular} \label{Tab:Betas+1-Loop}}
\end{table}

It is important to note here that all background configurations for the annulus topology of the worldsheet have been computed. For the 
M\"obius strip topology, however, the beta function coefficients only match the field theory expectation, or more concretely the explicit construction of open string states and associated Chan-Paton matrices, for D6-branes parallel along some two-torus $T^2_{(i)}$ if
$b_i\sigma_i\tau_i=0$, i.e. either the torus is of rectangular shape or some of the open string moduli vacuum expectation values ({\it vev}s,  
here displacements and Wilson lines $\sigma_i, \tau_i \in \{0,1\}$ for {\it fractional} and {\it rigid} D6-branes)
vanish, as first noted in Ref.~\citen{Forste:2010gw} and verified in the models constructed afterwards in Refs.~\citen{Honecker:2012qr,Ecker:2014hma,Ecker:2015vea}.
Since phenomenologically interesting models such as the MSSM example\cite{Ecker:2015vea} in table~\ref{Tab:D6-Config-MSSM} 
(which has $b_1=0$, $b_2=b_3=\nicefrac{1}{2}$) 
notoriously require
$b_i\sigma_i\tau_i \neq 0$, it is of great importance to compute this missing piece of information on the one-loop gauge threshold. 
In section~\ref{S:1-Loop}, we will come back to this issue and discuss for which values of K\"ahler moduli the missing information can be neglected or will even cancel out among various contributions.

Finally, the one-loop gauge threshold can be decomposed into contributions to the holomorphic gauge kinetic function and leading-order expressions for the open string K\"ahler metrics by matching with the standard SUGRA expressions.\cite{Akerblom:2007uc,Blumenhagen:2007ip,Honecker:2011sm}
At this point, it is particularly important for scalar potentials containing the shift-symmetric terms\cite{Hebecker:2012qp} $(H_u\overline{H}_d+h.c.)$
that all tree level K\"ahler metrics on toroidal orbifold background of Type IIA string theory  were found to be diagonal. It remains to be seen if this finding is specific to the simple backgrounds investigated so far, or if it generalizes to generic Calabi-Yau threefolds.

{\bf Yukawa and higher $m$-point functions:}
Each Yukawa coupling  consists of a product of a (holomorphic) classical worldsheet instanton sum,
which scales as $Y_{abc} \propto e^{-\sum_i \text{Area}^{(i)}_{abc}}$ in terms of the area bounded by three D6-branes $a,b,c$ with matter fields localised at the pairwise intersection points as computed in Refs.~\citen{Cremades:2003qj} for a factorisable six-torus $\prod_{i=1}^3 T^2_{(i)}$, and a (non-holomorphic) quantum prefactor 
that was derived in Refs.~\citen{Cvetic:2003ch,Abel:2003vv,Abel:2003yx,Lust:2004cx} again for the factorisable six-torus.
These results are valid  for {\it bulk} D6-branes (or fractions of pure {\it bulk} D6-branes on the $T^6/(\mathbb{Z}_2 \times \mathbb{Z}_{2M} \times \Omega {\cal R})$
orientifolds without discrete torsion e.g. in Refs.~\citen{Cvetic:2001tj,Cvetic:2001nr,Honecker:2003vq,Honecker:2003vw,Honecker:2004np,Forste:2010gw}) with Yukawa couplings arising at non-vanishing angles on all three tori.\footnote{Yukawa couplings on non-factorizable tori were recently considered in Refs.~\citen{Forste:2014bfa,Pesando:2015fpj}.}
The phenomenologically appealing models with a reduced or vanishing number of open string moduli, however, require 
{\it fractional}\cite{Blumenhagen:2002gw,Honecker:2004kb,Bailin:2006zf,Bailin:2007va,Gmeiner:2007zz,Bailin:2008xx,Gmeiner:2008xq,Gmeiner:2009fb,Honecker:2012jd,Honecker:2012fn,Bailin:2013sya,Seifert:2015fwr} or {\it rigid}\cite{Blumenhagen:2005tn,Forste:2010gw,Honecker:2012qr,Honecker:2013kda,Ecker:2014hma,Ecker:2015vea,Honecker:2015qba} D6-branes, respectively.
In contrast to the simple torus models, chiral matter can here also arise at intersections with one vanishing angle, e.g. $(0,\phi,-\phi)$, since
the $\mathbb{Z}_2$ symmetries break the naive ${\cal N}=2$ SUSY to ${\cal N}=1$ only. The argument that the CFT computation will give a vanishing result does thus not hold. 

Alternatively, it is known\cite{Cremmer:1982en,Dixon:1989fj,Cvetic:2003ch,Abel:2003yx,Lust:2004cx} that the quantum prefactor contains the K\"ahler metrics, $Y_{abc} \propto (K_{ab}K_{bc}K_{ca})^{-\nicefrac{1}{2}} e^{\kappa_4^2 {\cal K}/2}$ with  ${\cal K}$ the K\"ahler potential, 
of the relevant matter fields, which were determined to leading order as a byproduct of the gauge threshold computation in table~\ref{Tab:Betas+1-Loop}.
If the different particle generations arise from intersections at different angles, this non-holomorphic prefactor can lead to an additional mild hierarchy\cite{Honecker:2012jd,Honecker:2012fn} beyond the exponentials of worldsheet areas.

In the MSSM example of section~\ref{Sss:MSSM-Spectrum}, the relative displacement $\sigma_3^{ab}=1$ between the D6-brane stacks $a$ and $b$ 
 along $T^2_{(3)}$  in table~\ref{Tab:D6-Config-MSSM}  enforces a non-vanishing worldsheet area with the smallest suppression factors of trilinear 
 Yukawa couplings $Y_u \propto e^{-\frac{4v_2+v_3}{48}}$ and $e^{-\frac{v_2+4v_3}{48}}$ involving the D6-brane stacks $a$, $b$ and $d'$ and
$Y_d \propto e^{-\frac{v_3}{48}}$ involving the stacks $a$, $b$ and $c$, since the $d_R$ particles stem from $ac$ sectors while the $u_R$ states 
arise from $ad'$ sectors \cite{Ecker:2015vea}, cf. table~\ref{Tab:Spectrum-MSSM}.
Since all $ay$ sectors with $y \in \{b,c,c',d,d'\}$ intersect at angles $\pi(0,\phi,-\phi)$ with $\phi=\pm \frac{1}{6},\frac{1}{2}$, the quark K\"ahler metrics
are universal, $K_{Q_L}=K_{u_R}=K_{d_R}= e^{\phi_4}\sqrt{\frac{R_1^{(1)}}{R_2^{(2)}}\frac{1}{v_2v_3}} =K_{H_u}=K_{H_d}$,
and agree with the K\"ahler metrics of the Higgs fields arising also at angles $\pi(0,\phi,-\phi)$, but now with $\phi=\pm \frac{1}{3}$
in the $bc$ and $bd$ sectors.

Let us stress at this point that the full expression for Yukawa and higher $m$-point couplings at intersections of {\it fractional} or {\it rigid} D6-branes has
to our best knowledge never been computed, and complex phases might arise when correctly defining the relevant boundary vertex operators.
Such phases would be of major relevance to phenomenology. 
One possibility to check consistency of a first computation of this type would be to reproduce the one-loop gauge thresholds described above. 
This might also furnish the missing results for the M\"obius strip topology with $b_i\sigma_i\tau_i \neq 0$.

{\bf K\"ahler metrics at one-loop and D-brane instantons:}
The above CFT results can be extended in two ways: 
at the perturbative level, in Refs.~\citen{Berg:2011ij,Berg:2014ama} the first computations of one-loop corrections to the 
K\"ahler metrics in IIA orientifold models were performed. The computations and resulting non-renormalization are, however, 
only valid for {\it bulk} D6-branes, due to the prefactors ${\rm tr}\gamma_{\mathbb{Z}_2}=0$ of the annulus amplitudes used there.
All string vacua with chiral matter on {\it fractional}\cite{Blumenhagen:2002gw,Honecker:2004kb,Bailin:2006zf,Bailin:2007va,Gmeiner:2007zz,Bailin:2008xx,Gmeiner:2008xq,Gmeiner:2009fb,Honecker:2012jd,Honecker:2012fn,Bailin:2013sya,Seifert:2015fwr} or {\it rigid}\cite{Blumenhagen:2005tn,Forste:2008ex,Forste:2010gw,Honecker:2012qr,Honecker:2013kda,Ecker:2014hma,Ecker:2015vea,Honecker:2015qba} D6-branes violate this condition, making the extension of these  computations to non-trivial twisted annulus contributions necessary.
The second generalization of the above CFT methods consists in considering non-perturbative effects from D2-brane instantons wrapped along compact three-cycles.
The computation of their contribution to the superpotential $\propto e^{-{\cal S}_{\text{inst}}}$ requires an integration over the D2-brane 
zero modes, which leads to a vanishing result unless the number of zero modes is minimal, see e.g. the review articles Refs.~\citen{Blumenhagen:2009qh,Cvetic:2011vz} for details. Non-perturbative corrections to the superpotential arise thus only from $O(1)$  D2-branes wrapping {\it rigid} three-cycles.
While these contributions have so far not been computed for the MSSM example of section~\ref{Sss:MSSM-Spectrum}, 
the classification of three-cycles wrapped by probe D6-branes with $USp(2)$ gauge group in Ref.~\citen{Ecker:2015vea} is identical to the classification of $O(1)$  D2-branes, and the explicit form of the non-perturbative superpotential is thus in principle within reach.
 
\vspace{2mm}

In summary, a plethora of partial results on  gauge and Yukawa couplings for D6-branes on toroidal orbifolds is known, but further intensive CFT computations are required to provide the exact formulas for gauge couplings and K\"ahler metrics beyond leading order, prefactors/phases of perturbative Yukawa couplings
as well as non-perturbative contributions to the superpotential.

\subsubsection{Moduli stabilization at the orbifold point}\label{Sss:ModuliStab}

Generic Type IIA orientifold compactifications contain in the closed string sector $h^{11}_-$ K\"ahler and $h^{21}$ complex structure moduli,
plus in the open string sector displacement and Wilson line moduli transforming in the adjoint representation of the gauge group.
The latter kind of moduli is absent by construction if one uses {\it rigid} three-cycles. Toroidal orbifold backgrounds containing a 
$\mathbb{Z}_2 \times \mathbb{Z}_2$ subgroup with non-trivial discrete torsion phase among the two $\mathbb{Z}_2$ factors provide 
such three-cycles for $T^6/(\mathbb{Z}_2 \times \mathbb{Z}_{2M})$ with $M \in \{1,3,3'\}$ \cite{Blumenhagen:2005tn,Forste:2010gw,Honecker:2012qr,Ecker:2014hma,Ecker:2015vea}, as exemplified above in section~\ref{Sss:MSSM-Spectrum}.
Additionally, in Ref.~\citen{Marchesano:2014iea} it was conjectured that open string moduli receive a non-trivial potential via backreaction on the geometry,
at least if closed string background fluxes are turned on, which, however, impedes model building with currently known geometric and CFT techniques.
Also the $T^6/(\mathbb{Z}_6' \times \Omega {\cal R})$ model in Ref.~\citen{Bailin:2011am} employs closed string background fluxes to stabilize geometric closed string moduli.

Since open and closed string sectors couple to each other, e.g. via the DBI and CS actions invoked in Eq.~\eqref{Eq:DBI}, it is natural to hypothesize that the presence of some D-brane - even without closed string fluxes - will stabilize (at least some of) the closed string moduli which it couples to. From a four-dimensional effective field theory point of view, the contribution to the scalar potential stems from a D-term, $V_{\text{scalar}}= \frac{1}{2} D_a^2 + \ldots$, 
in which the Fayet-Iliopoulos (FI) term $\zeta_i \supset D_a$  corresponds to the {\it vev} of some geometric modulus.
From a microscopic point of view, this hypothesis can be probed as follows: 
the volume of a {\it fractional} or {\it rigid} D6-brane at the singular orbifold point, where all twisted complex structure moduli have a vanishing {\it vev}, 
is given by the corresponding fraction of its bulk part. For SUSY D6-branes, the {\it sLag} condition in Eq.~\eqref{Eq:sLag} implies that $\int_{\Pi_a} \Omega_3 \stackrel{\text{SUSY}}{=}\text{Vol}(\text{D}6_a)$, in other words the period is real. In Refs.~\citen{Blaszczyk:2014xla,Blaszczyk:2015oia,Koltermann:2015oyv,Koltermann:2016} a hypersurface parameterization of the factorizable six-torus and its $\mathbb{Z}_2 (\times \mathbb{Z}_2)$ orbifolds\cite{Vafa:1994rv} was implemented
to allow for (complex structure) deformations away from the singular point. The {\it sLag} condition on the cycle $\Pi_a$
 is then probed by computing the corresponding period and verifying if it remains real or develops an imaginary part upon deformation.
By the relation in Eq.~\eqref{Eq:g-tree}, any change in the cycle volume also changes the gauge coupling of the associated D-brane stack. The following different cases arise:\cite{Blaszczyk:2014xla,Blaszczyk:2015oia,Koltermann:2015oyv,Koltermann:2016}
\begin{enumerate}
\item
If a D6-brane couples via some orientifold-odd exceptional three-cycle within $\Pi_a^{\mathbb{Z}_2}$ to a twisted complex structure modulus,
its deformation will break SUSY and generate a D-term potential. Deforming several singularities simultaneously leads to an additive scalar potential of the form $\propto \sum_i \zeta_i^2$. 
\item
If a D6-brane only couples via some orientifold-even exceptional three-cycle to a twisted complex structure modulus, its deformation
will change the value of the period, roughly speaking as $\pm \sqrt{\varepsilon}$ with $\varepsilon$ the SUSY deformation parameter 
and the sign factor depending on how the exceptional cycle enters $\Pi_a^{\mathbb{Z}_2}$. 
The gauge coupling of Eq.~\eqref{Eq:g-tree} thus experiences a flat direction.
\item
If a D6-brane does not couple directly to some twisted modulus, switching on a {\it vev} will only backreact on the D6-brane volume and gauge coupling 
via (gravitational) higher order effects. This statement holds for K\"ahler as well as complex structure moduli.
\end{enumerate}
In the MSSM example of table~\ref{Tab:D6-Config-MSSM}, the bulk complex structure modulus of $T^2_{(1)}$ and the three bulk K\"ahler moduli 
are not stabilized by their couplings to D-branes, and neither are the twelve twisted K\"ahler moduli at $\mathbb{Z}_6'$ and $\mathbb{Z}_3$
singularities. In the $\mathbb{Z}_2^{(1)}$ twisted sector, three twisted complex structure moduli are not coupled to any D6-brane, two are stabilized by couplings to D6-branes $a,c,h$ and one provides a flat direction in the gauge couplings of stacks $b$ and $d$.
While at the orbifold point, only {\it relative} $\mathbb{Z}_2 \times \mathbb{Z}_2$ eigenvalues are relevant, upon deformation the {\it absolute} sign 
becomes important as anticipated in item~(2). In any case, if the volume of brane $b$ shrinks, the volume of brane $d$ will increase, or vice versa. 
In the $\mathbb{Z}_2^{(2)}$ twisted sector, branes $c$ and $d$ couple to two of the four twisted deformation moduli, 
and in the $\mathbb{Z}_2^{(3)}$ twisted sector branes $a$, $d$ and $h$ do the same. 
Overall, we thus expect six twisted deformation moduli to be stabilized at the singular orbifold point by their couplings to the five stacks of D6-branes;
one twisted deformation modulus constitutes a flat direction, which affects the $SU(2)_W$ and the $U(1)_Y$ gauge coupling strengths; finally
the remaining deformation moduli possess flat directions, but backreact on the geometry via (gravitational) higher order effects.\cite{Honecker:2016-3}

In principle, a FI term could be compensated by a suitable {\it vev} of some open string scalar. 
Giving a scalar partner of some MSSM fermion a {\it vev} would result in breaking the Standard Model gauge group, leaving 
only the option of giving a {\it vev} to the right-handed sneutrino in table~\ref{Tab:Spectrum-MSSM}. But such a {\it vev} can at most
compensate D-terms of either brane $c$ or $d$ due to the opposite $U(1)_c \times U(1)_d$ charges.
Similarly, a {\it vev} of some right-handed squark would generically produce a D-term for stack $c$ or $d$ while eliminating that of stack $a$.

\vspace{2mm}

In summary, the mere existence of D-branes creates a backreaction on the geometry which leads to the stabilization of some geometric moduli, 
even before invoking closed string fluxes. The appeal of this mechanism lies in the fact that it avoids the no-go statement of simultaneously closed string fluxes and chiral fermions localized on some cycle.\cite{Villadoro:2006ia,Blumenhagen:2007sm}
Of course, it might be attractive to switch on closed string fluxes and stabilize from the particle physics sector decoupled moduli, e.g. in view of viable inflationary potentials.

\section{Heterotic String Theories and Non-Perturbative Regimes}\label{S:Het}

In section~\ref{S:Pert-II}, all allowed matter representations and corresponding gauge groups from geometrically engineered
perturbative Type II string theory were discussed.
Here, a complementary view on the allowed maximal rank of the overall gauge group by means of S-duality to the heterotic $SO(32)$ string theory will be discussed
as well as  enhancements of the allowed set of representations and simple Lie groups in the non-perturbative regime of Type II string theory, so-called {\it F-theory}, 
or the heterotic $E_8 \times E_8$ string theory. Finally, we briefly discuss further enhancements due to singularities within the compact six-dimensional space.

\subsection{Heterotic $SO(32)$ on Calabi-Yau manifoldss}\label{Ss:SO32}

On non-singular Calabi-Yau threefolds, Type IIA orientifolds with intersecting D6-branes have been conjectured to be mirror dual to Type IIB orientifolds with 
magnetized D7/D3- or D9/D5-brane systems, which in turn are conjectured to be S-dual to compactifications of the $SO(32)$ heterotic string theory. 
In the latter framework, the appearance of bifundamental representations can be seen from the embedding of $U(n_i)$ vector bundles $V_i$ within the ten-dimensional gauge group
$SO(32)$, which leads to the breaking of the perturbative gauge group to $\prod_{i=1}^K U(N_in_i) \times SO(2M) \to \prod_{i=1}^K U(N_i) \times U(n_i) \times SO(2M)$
and all matter states arising from the decomposition of the adjoint representation of $SO(32)$:\cite{Blumenhagen:2005pm,Blumenhagen:2005zg,Honecker:2006dt}
\begin{equation}\label{Eq:496-decompose}
{\bf 496} \to \left(\begin{array}{c} ({\bf Anti}_{SO(2M)}) + \sum_{i=1}^K ({\bf Adj}_{U(N_i)};{\bf Adj}_{U(n_i)}) \\
 \sum_{i=1}^K [ ({\bf Anti}_{U(N_i)};{\bf Sym}_{U(n_i)}) + ({\bf Sym}_{U(N_i)};{\bf Anti}_{U(n_i)}) + h.c. ] \\
 \sum_{i<j} [ ({\bf N}_i,{\bf N}_j;{\bf n}_i, {\bf n}_j) + ({\bf N}_i,\overline{\bf N}_j;{\bf n}_i, \overline{\bf n}_j) + h.c.]
 +  \sum_{i=1}^K [ ({\bf 2M},{\bf N}_i, {\bf n}_i) + h.c. ]
\end{array}\right)
.
\end{equation}
Non-perturbative five-branes in the $SO(32)$ heterotic string theory wrapped on compact two-cycles $\Gamma_j$ (Poincar\'{e} dual to the four-forms $\gamma_j$)
support skyscaper sheaves and lead to the four-dimensional  gauge group $USp(2M)$ with bifundamental matter arising with $U(N_i)$ gauge factors 
as well as further $USp(2M)$'s.
The massless matter spectrum is counted in terms of dimensions of cohomology and extension groups, and the net-chirality is determined by the corresponding Euler characteristic,
\begin{equation}\label{Eq:EulerChar}
\chi (W) = \int_{CY_3} \left({\rm ch}_3(W) + \frac{1}{12} c_2(T) c_1(W) \right)
\quad \text{ or } \quad
-  \int_{CY_3} c_1(V_i) \wedge \gamma_j
,
\end{equation}
where ${\rm ch}_n(W)$ denotes the $n^{\text{th}}$ Chern character of the vector bundle $W$, $c_1(W)$ its first Chern class and $c_2(T)$ the 
second Chern class of the tangent bundle to the manifold.
The chirality of the bifundamental representation $({\bf N}_i,\overline{\bf N}_j)$ is for example computed using the bundle $W = V_i \otimes V_j^{\ast}$,
while the second expression in Eq.~\eqref{Eq:EulerChar} is valid for bifundamental matter arising from a vector bundle $V_i$ in combination with a five-brane
wrapped on the two-cycle $\Gamma_j$.

The necessary string theoretic consistency conditions consist of the Bianchi identity on the three-form field strength, $dH_3=0$,
\begin{equation}\label{Eq:Bianchi}
\sum_i N_i {\rm ch}_2 (V_i) - \sum_j M_j \gamma_j = - c_2(T)
.
\end{equation}
The existence of well-defined spinors on the Calabi-Yau manifold - or in other words the absence of a global Witten anomaly - is guaranteed by   
the following `mod 2' condition:
\begin{equation}\label{Eq:het-spinors}
\sum_i N_i c_1(V_i)  \in H^2(CY_3, 2 \mathbb{Z})
.
\end{equation}
When reducing to rank $n_i \equiv 1$ for all $i$, one can see that via S-duality and mirror symmetry the Bianchi identity in Eq.~\eqref{Eq:Bianchi}
corresponds to the RR tadpole cancellation condition in Eq.~\eqref{Eq:RRtcc}, and the constraint on the existence of well-defined spinors
in Eq.~\eqref{Eq:het-spinors} is mapped to the K-theory constraint in Eq.~\eqref{Eq:Kconstr}.

While in Type II string theories, the Green-Schwarz anomaly cancellation terms arise from the leading order of the DBI and CS actions,
in compactifications of heterotic string theories, the  counterterm arises at one-loop in the ten-dimensional supergravity action.\cite{Green:1984sg}

The dual viewpoint of heterotic string theory provides an intuitive insight into the stringent upper bound on the rank of the total gauge group, 
since here the perturbative gauge group is in embedded into $SO(32)$ (or $E_8 \times E_8$ as discussed in the next section~\ref{Ss:E8+F}), 
and any additional non-perturbative $USp(2M)$ gauge factor is constrained by the Bianchi identity in Eq.~\eqref{Eq:Bianchi}.

\subsection{Heterotic $E_8 \times E_8$ and F-theory  on Calabi-Yau manifolds}\label{Ss:E8+F}

The discussion of the Bianchi identity, well-definedness of spinors and net-chirality is completely analogous for the $SO(32)$ and $E_8 \times E_8$ case.\cite{Blumenhagen:2005ga,Honecker:2006dt}
Also for compactifications of the $E_8 \times E_8$ heterotic string, simple considerations of representation theory severely constrain any new physics
state.
Considering for example the series of decompositions
$E_8^{(i)} \to E_{r_i} \times SU(n_i+m_i) \to E_{r_i} \times SU(n_i) \times SU(m_i) \times U(1)$ with 
$E_{r_i} \in \{E_7, E_6, SO(10), SU(5), SU(3) \times SU(2) \}$ and $r_i+n_i+m_i=9$
leads in the first step to the following decompositions of the adjoint representation of $E_8$:
\begin{equation}\label{Eq:248-decompose}
{\bf 248} \to \left\{
\begin{array}{cc} 
({\bf 133},{\bf 1}) + ({\bf 1},{\bf 3}) + ({\bf 56},{\bf 2}) & {E_7 \times SU(2)}
\\
 ({\bf 78},{\bf 1}) + ({\bf 1},{\bf 8}) + [ ({\bf 27},{\bf 3}) + h.c.]  & {E_6 \times SU(3)} 
\\
({\bf 45},{\bf 1}) + ({\bf 1},{\bf 15}) + ({\bf 10}, {\bf 6}) +  [({\bf 16},{\bf 4})  + h.c.] & {SO(10) \times SU(4)}
\\
({\bf 24},{\bf 1}) + ({\bf 1},{\bf 24}) + [ ({\bf 5},\overline{\bf 10}) + ({\bf 10},{\bf 5})  + h.c.] & {SU(5) \times SU(5)}
\\
\left.
\begin{array}{c} ({\bf 8},{\bf 1};{\bf 1}) + ({\bf 1},{\bf 3};{\bf 1}) + ({\bf 1},{\bf 1};{\bf 35}) \\
+ ({\bf 1},{\bf 2};{\bf 20}) + [ ({\bf 3},{\bf 2};{\bf 6}) + (\overline{\bf 3},{\bf 1};{\bf 15}) + h.c.]
\end{array}\right\}
 & {SU(3) \times SU(2) \times SU(6)}
\end{array} \right.
.
\end{equation}
Besides the exceptional GUT group $E_6$, now also the spinor representation $({\bf 16})$ of $SO(10)$ appears. 
It is also noteworthy that the only apparent representations of the $SU(n+m)$ factors in Eq.~\eqref{Eq:248-decompose} are the fundamental $({\bf n+m})$, antisymmetrics $(\frac{\bf (n+m)(n+m-1)}{\bf 2})$ and $(\frac{\bf (n+m)(n+m-1)(n+m-2)}{\bf 6})$, and the adjoint $({\bf(n+m)^2-1})$.
This finding again poses constraints on the possible appearance of states with exotic charges.
Let us for concreteness focus on the last case with $SU(n+m)=SU(6)$ and perform 
 the second step of decomposing $SU(6) \to SU(n) \times SU(6-n) \times U(1)$ 
as exemplified in table~\ref{Tab:SU6-embeddings}.
\begin{table}[ht!]
\tbl{Embedding of $SU(n) \times SU(6-n)  \times U(1) \subset SU(6)$ bundles.}
{\begin{tabular}{@{}|c|c||c|c|@{}}\toprule
 $SU(6)$ rep. & $SU(5) \times U(1)$ & $SU(4) \times SU(2) \times U(1)$ & $SU(3) \times SU(3) \times U(1)$ 
\\\colrule
 $({\bf 6})$ & $({\bf 5})_1 + ({\bf 1})_{-5}$  & $({\bf 4},{\bf 1})_1 + ({\bf 1},{\bf 2})_{-2}$ & $({\bf 3},{\bf 1})_1 + ({\bf 1},{\bf 3})_{-1}$
\\
 $({\bf 15})$ & $({\bf 10})_2 + ({\bf 5})_{-4}$ & $({\bf 6},{\bf 1})_2 + ({\bf 1},{\bf 1})_{-4}+ ({\bf 4},{\bf 2})_{-1}$
& $(\overline{\bf 3},{\bf 1})_2 + ({\bf 1},\overline{\bf 3})_{-2} + ({\bf 3},{\bf 3})_0$
\\
 $({\bf 20})$ & $({\bf 10})_{-3}+ (\overline{\bf 10})_3$ &  $({\bf 4},{\bf 1})_{-3} + (\overline{\bf 4},{\bf 1})_3 + ({\bf 6},{\bf 2})_0$
& $({\bf 1},{\bf 1})_{-3} + ({\bf 1},{\bf 1})_3 + ({\bf 3},\overline{\bf 3})_{-1} +  (\overline{\bf 3},{\bf 3})_1$
\\
  $({\bf 35})$ & $\begin{array}{c} ({\bf 1})_0 + ({\bf 5})_6 + (\overline{\bf 5})_{-6} \\ + ({\bf 24})_0 \end{array}$
& $\begin{array}{c} ({\bf 1},{\bf 1})_0 + ({\bf 15},{\bf 1})_0 + ({\bf 1},{\bf 3})_0 \\  + ({\bf 4},{\bf 2})_3  + (\overline{\bf 4},{\bf 2})_{-3} \end{array}$
& $\begin{array}{c} ({\bf 1},{\bf 1})_0+ ({\bf 8},{\bf 1})_0 + ({\bf 1},{\bf 8})_0 \\  + ({\bf 3},\overline{\bf 3})_2  + (\overline{\bf 3},{\bf 3})_{-2} \end{array}$
\\\botrule
\end{tabular} \label{Tab:SU6-embeddings}}
\end{table}
While one might naively try to embed only a $SU(n)$ bundle $V_n$ and  identify $U(1)_Y \simeq \frac{1}{6}U(1)$ in table~\ref{Tab:SU6-embeddings}, 
this will generically lead to an enhancement of the four-dimensional gauge group.
E.g. for $n=5$, both left-handed quarks $Q_L$ and right-handed up-type quarks $u_R$ are associated to the bundle $V_5$, while left-handed leptons $L$ and right-handed down-type quarks $d_R$ are linked to the bundle $\wedge^2 V_5$, leading to an $SU(5)$ GUT instead of $SU(3) \times SU(2) \times U(1)_Y$.
In Refs.~\citen{Blumenhagen:2005ga,Blumenhagen:2006ux,Blumenhagen:2006wj}, therefore a program to embed $U(n) \times U(m)$ bundles such that the hypercharge remains massless was initiated, with a plethora of systematic
model searches using only line bundles conducted afterwards by various groups.\cite{Blumenhagen:2010pv,Anderson:2011ns,Anderson:2012yf,Buchbinder:2013dna,Anderson:2013xka,Nibbelink:2015ixa} The advantage of such line bundle constructions consists
in the fact that line bundles are by definition stable.

While five-branes in compactifications of the $E_8 \times E_8$ heterotic string theory contribute to the Bianchi identity analogously to Eq.~\eqref{Eq:Bianchi}, 
in contrast to the $SO(32)$ case each five-brane supports a tensor multiplet in six dimensions, which reduces to an Abelian gauge multiplet in four dimensions.
Due to the very different origin of these $U(1)$ factors,  matter fields remain uncharged.

\vspace{2mm}

The message to take away here is that again the origin of all charged states from the adjoint representation(s) of the ten-dimensional gauge group $E_8 (\times E_8)$ severely constrains the representations under which any new physics particle and/or group might transform; for a 
comprehensive list on the relevant branchings of representations see e.g. the report in Ref.~\citen{Slansky:1981yr}.

\vspace{2mm}

{\bf F-theory} is defined as the strong coupling limit of Type IIB string theory with D7/D3-brane systems,
where the string coupling is identified with the complex structure of an auxiliary two-torus or fibration. In this picture,
Kodaira's ADE-classification of singularities (along the fiber) provides a geometric realisation of the non-Abelian gauge groups
with sections corresponding to the Abelian factors\cite{Morrison:1996na,Morrison:1996pp}, for an extended overview see e.g. the lecture notes in Ref.~\citen{Weigand:2010wm}.
The revival of four-dimensional F-theory models in 2008\cite{Donagi:2008ca,Beasley:2008dc} has consolidated  
the geometric brane engineering with the $E_8 \times E_8$ heterotic dual (whenever existent) construction using vector bundles.
Attempts of classifying all possible SCFTs in six dimensions have recently been made\cite{Heckman:2013pva,Heckman:2015bfa}.
While the question, how vast the four-dimensional landscape of string and F-theory vacua is, remains open, it is again obvious that 
new physics states are constrained by representation theory of non-Abelian gauge groups, which in the F-theory language corresponds
to the classification of singularities.

Field theoretical investigations of both heterotic string theories on smooth manifolds 
are constrained by the limited scope of dimensionally reducing the ten-dimensional SUGRA action, 
just as for Type II orientifolds on smooth backgrounds. 
For perturbative Yukawa couplings in heterotic string theories, bundle cohomologies can be invoked to compute the holomorphic factors~\cite{Anderson:2009ge,Anderson:2010tc}. Considerations on the existence of non-vanishing non-perturbative contributions to the superpotential 
can e.g. be found in Refs.~\citen{Braun:2007tp,Braun:2007xh,Braun:2007vy} in relation to Gromov-Witten invariant of the underlying Calabi-Yau geometry.
In the non-perturbative F-theory corner, in addition to the limitations on the effective action of the 
perturbative $E_8 \times E_8$ heterotic string theory
 it remains to be clarified to what extent the perturbative SUGRA terms compete with e.g. non-perturvative  instanton corrections.

\subsection{Heterotic on Toroidal Orbifolds}\label{Ss:HetOrbifolds}

In contrast to the toroidal orbifolds of Type II string theories discussed in section~\ref{S:Pert-II}, 
in compactifications of heterotic string theory both geometric moduli and matter fields arise from excitations of closed strings.
In the latter case, slight enhancements of representations and/or gauge groups at orbifold singularities occur. The 
characteristic features can already be observed in the overview of heterotic orbifold spectra in six dimensions in Ref.~\citen{Honecker:2006qz}:
the Bianchi identity and spinorial condition in Eqs.~\eqref{Eq:Bianchi} and~\eqref{Eq:het-spinors} translate 
into a quadratic and a linear condition on the shift vectors $\vec{V}$ and $\vec{v}$ in the gauge and torus lattice, respectively:
\begin{equation} 
N \left( \sum_k V_k^2 - \sum_i v_i^2\right) = 0 \text{ mod }2
\quad \text{and} \quad
N \sum_k V_k= 0 \text{ mod }2,
\end{equation}
for a perturbative $T^{2n, n \in\{2,3\}}/\mathbb{Z}_N$ compactification.
The total rank for any compactification without non-perturbative heterotic five-branes amounts to 16, with 
either some additional massless Abelian gauge factor or a slightly enhanced non-Abelian gauge group.
At orbifold singularities, twisted closed strings can combine into slightly enhanced representations, with in particular
the spinor representation appearing in $SO(32)$ heterotic orbifold compactifications.
If the gauge group contains $U(1)$ factors, charges in the $k^{\text{th}} $twisted sector are typically shifted by $\nicefrac{k}{N}$
compared to those expected from the decompositions in Eqs.~\eqref{Eq:496-decompose} and \eqref{Eq:248-decompose}.

The naive connection to heterotic compactifications on manifolds with vector bundles
consists of identifying the shift vector with some line bundle,\cite{Honecker:2006qz,Honecker:2007uw}
\begin{equation}
\frac{1}{N}(1_{n_1},2_{n_2}, \ldots,0_{n_0}) \to (L_{n_1},L_{n_2}^2,\ldots,{\cal O}),
\end{equation}
but within four-dimensional models, matter representations at singularities often do not allow for a unique 
identification of the twisted geometric modulus, admitting instead for various resolutions to different Calabi-Yau manifolds.\cite{Blaszczyk:2010db,NibbelinkGroot:2010wm,Blaszczyk:2011hs}

CFT methods to compute couplings for heterotic orbifold models were recently reconsidered\cite{Nilles:2011aj,Kobayashi:2011cw}
due to the extended scans of landscape patches and studies of their phenomenological features.\cite{Lebedev:2006kn,Lebedev:2008un}

Free fermionic models have recently been shown to be dual\cite{Donagi:2008xy,Athanasopoulos:2016aws} to $T^6/(\mathbb{Z}_2 \times \mathbb{Z}_2)$ orbifolds in the bosonic formulation as classified in Refs.~\citen{Donagi:2008xy,Fischer:2012qj}.

\subsection{Gepner/RCFT models}\label{Ss:Gepner}

The classifications of particle spectra in Gepner or RCFT models\cite{Dijkstra:2004ym,Dijkstra:2004cc}
belong morally to non-perturbative string theory corners away from the geometric engineering regime
presented in section~\ref{S:Pert-II}. Similarly to F-theory or heterotic orbifold models, enhancements of 
gauge groups occur. As a concrete example, hypercharge embeddings have been classified\cite{Anastasopoulos:2006da},
adding the values $x=-\frac{1}{2},1,\frac{3}{2}$ in
\begin{equation}
(q_a,q_b,q_c,q_d) \in \left\{ \left(\frac{-1}{3},\frac{-1}{2},0,0 \right),
 \left(\frac{1}{6},\frac{1}{2},0,\frac{-3}{2} \right) \text{ and }
\left(x-\frac{1}{3},x-\frac{1}{2},x,1-x \right) \text{ with } x=-\frac{1}{2},0,\frac{1}{2},1,\frac{3}{2} \right\}
\end{equation}
to the classification from a gauge quiver point of view in Eq.~\eqref{Eq:possible-Ys}  of section~\ref{Sss:Zn}.
The extensive list of RCFT vacua thus again provides only a rather small number of possibilities to obtain the Standard Model gauge group.

\section{Open String Axions}\label{S:Axions}

In 1977, the QCD axion was proposed as the Goldstone mode of a spontaneously broken global $U(1)_{PQ}$ symmetry~\cite{Peccei:1977hh,Peccei:1977ur}.
The natural embedding into Type II string theory consists in identifying the $U(1)_{PQ}$ symmetry with a massive gauge symmetry 
that acts as a global symmetry in perturbation theory, but is broken by non-perturbative effects such as D-brane instantons as discussed in section~\ref{Sss:Zn}.
Depending on the specific string vacuum, a discrete $\mathbb{Z}_n$ subgroup
such as the $\mathbb{Z}_3$ gauge symmetry in the MSSM example of section~\ref{Sss:MSSM-Spectrum} can remain as exact symmetry of the low-energy effective field theory.
The QCD axion then arises as a massless open string state\cite{Kiritsis:2003mc,Svrcek:2006yi,Berenstein:2012eg,Honecker:2013mya,Honecker:2015ela} with $\mathbb{Z}_n \subset U(1)_{PQ}$ charge.

Since string theory vacua are  customarily constructed to preserve ${\cal N}=1$ SUSY and thus have at least on pair of $(H_u,H_d)$ Higgs doublets,
it is natural to extend the DFSZ axion model first proposed in 1981.~\cite{Zhitnitsky:1980tq,Dine:1981rt} In order to do so, the following set of constraints has to be met:
\begin{itemize}
\item
The QCD axion $\sigma$ is charged under the massive $U(1)_{PQ}$ symmetry, but transforms as a singlet under the Standard Model gauge group.
\item
The Higgses $(H_u,H_d)$ also possess $U(1)_{PQ}$ charges in order to allow for a non-trivial Higgs-axion potential.
\item
To allow for all Yukawa couplings, either right- or left-handed quarks and leptons possess $U(1)_{PQ}$ charges.
\end{itemize}
While from a low-energy perspective, a shift by the hypercharge can be used to ensure that either left- or right-handed quarks (and similarly leptons) are neutral under $U(1)_{PQ}$, it is also useful to demand that the mass eigenstates of the massive $U(1)_{PQ}$ and massless $U(1)_Y$ symmetries are orthogonal in Type II string theory models.
Based on these constraints, there exist only two possible realizations of $U(1)_{PQ}$ and the associated QCD axion for the `standard' hypercharge embedding:\cite{Honecker:2013mya,Honecker:2015ela}
\begin{enumerate}
\item $U(1)_{PQ} \simeq U(1)_b \subset U(2)_b$: in this case, the QCD axion is realized as the antisymmetric representation (or its conjugate)
 of $U(2)_b$ and has thus the $U(1)_{PQ}$ charge $\pm 2$. The left-handed Standard Model particles carry $U(1)_{PQ}$ charge, while the right-handed quarks and leptons are neutral.
\item $U(1)_{PQ} \simeq U(1)_c - U(1)_d$: in this case, the QCD axion is the pseudoscalar partner of some right-handed neutrino transforming as $({\bf 1})_{\pm 1,\mp 1}$ under $U(1)_c \times U(1)_d$ and thus again with charge $\pm 2$ under $U(1)_{PQ}$. The right-handed Standard Model particles carry $U(1)_{PQ}$ charge, whereas the left-handed ones remain neutral. 
\end{enumerate}
In both realizations, charge selection rules allow for the following Higgs-axion potential,\cite{Honecker:2013mya}
\begin{eqnarray}
V_{\text{DFSZ}}^{\text{stringy}}=& V_D + V_F + V_{\text{soft}}
\nonumber\\
=& \lambda_u (H_u^{\dagger}H_u - v_u^2)^2 + \lambda_d (H_d^{\dagger}H_d - v_d^2)^2 + \lambda_{ud} |H_u^{\dagger}H_d|^2
+ \lambda_{\sigma} (\sigma^{\dagger}\sigma - v_{\sigma}^2)^2
\nonumber\\
& + (\tilde{\lambda}_{u\sigma} H_u^{\dagger}H_u + \tilde{\lambda}_{d\sigma} H_d^{\dagger}H_d)\sigma^{\dagger}\sigma
+ \tilde{\lambda}_{ud} |H_u \cdot H_d|^2 
\nonumber\\
& + \widehat{\lambda}_{ud\sigma} (H_u \cdot H_d \sigma + h.c.) 
,
\label{Eq:DFSZpotential}
\end{eqnarray}
which differs from the original DFSZ model in the last line, $V_{\text{DFSZ}}^{\text{original}} \supset H_u \cdot H_d \sigma^2$,
due to the constraint that the complex scalar $\sigma$ obtains it $U(1)_{PQ}$ charge from the endpoints of some open string 
and thus has twice the charge of the field theoretically predicted model. As a consequence the Higgs-axion coupling $\widehat{\lambda}_{ud\sigma}$
on the last line of Eq.~\eqref{Eq:DFSZpotential} does not experience the $\nicefrac{1}{M_{\text{Planck}}}$ suppression of
the original DFSZ model.

\vspace{2mm}

The soft breaking terms in Eq.~\eqref{Eq:DFSZpotential} are expected to arise upon spontaneous SUSY breaking
in a hidden sector, which is then via couplings to the gravity multiplet transmitted to the Higgs-axion sector. 
For example, in the MSSM-like model of section~\ref{Sss:MSSM-Spectrum} a gaugino condensate of the $SU(4)_h$ group might form 
if the associated vector-like matter states listed in table~\ref{Tab:Spectrum-MSSM} decouple at a sufficiently high energy scale. 
On the other hand, mass terms for the vector-like down-type quarks arise via three-point couplings to the axion multiplets $\Sigma$ 
with a somewhat weaker suppression of e.g. $e^{-\frac{v_3}{12}}$ than the Yukawa couplings discussed in section~\ref{Sss:BLO}.
Balancing all phenomenological constraints is therefore expected to require a considerable fine-tuning of Higgs and saxion {\it vev}s.

\section{One-Loop Effects and Massive Gauge Bosons}\label{S:1-Loop}

Balancing the string scale $M_{\text{string}}$, string coupling $g_{\text{string}}$ and Calabi-Yau and three-cycle volumes $\text{Vol}(CY_3)$ and $\text{Vol}(\Pi_x)$, respectively, such that gravity in four dimensions is weak, but the gauge couplings sufficiently strong, can in principle
be achieved in two different ways:
\begin{enumerate}
\item
 The tree-level gauge couplings in Eq.~\eqref{Eq:g-tree} can be rewritten as,
 \begin{equation}\label{Eq:g-tree-rewritten}
 \frac{4\pi}{g_{a,\text{tree}}^2} = \frac{1}{2\sqrt{2}c_ak_a}
 \frac{M_{\text{Planck}}}{M_{\text{string}}} \prod_{i=1}^2 \sqrt{V_{aa}^{(i)}}
 ,
 \end{equation}
 with the square root of the left hand side of Eq.~\eqref{Eq:M-Pl-M-str} appearing explicitly here.
 $V_{aa}^{(i)}$ has been defined in table~\ref{Tab:Betas+1-Loop} and only depends on toroidal wrapping numbers and complex structure moduli
 via $(R_1^{(i)}/R_2^{(i)})^{\pm 1}$. For either wrapping numbers $n^i_a=0$ or \mbox{$m^i_a+b_in^i_a=0$}, a large hierarchy among the two radii can thus compensate 
 a large ratio of $M_{\text{Planck}}/M_{\text{string}}$ in Eq.~\eqref{Eq:g-tree-rewritten}.

Type IIA orientifold compactifications on toroidal orbifolds usually require balancing RR charges among O6-planes extended along both
$n^i_a=0$ and $m^i_a+b_in^i_a=0$, which impedes large hierarchies among the associated radii and thus makes it natural to assum gauge coupling unification at $M_{\text{string}} \sim M_{\text{GUT}} = {\cal O}(10^{16})$ GeV\cite{Blumenhagen:2003jy}. 

The MSSM example in table~\ref{Tab:D6-Config-MSSM}, however, has the special feature that $m^1_x=0=b_1$ for all D6-brane stacks.
On this particular orbifold background with discrete torsion, the O6-plane charge of the $\Omega{\cal R}\mathbb{Z}_2^{(2)}$-invariant plane is
cancelled by the exotic charge of the  $\Omega{\cal R}\mathbb{Z}_2^{(3)}$-invariant plane along $T^2_{(1)}$.
 A very unisotropic compactification with e.g. $R_2^{(1)} \sim 10^{7} R_1^{(1)}$ can thus compensate for a lower value of the string scale
 $M_{\text{string}} \sim 10^{12}$ GeV.
 The evaluation of the relevant length scales in the MSSM example of section~\ref{Sss:MSSM-Spectrum}
 further gives the relation $\nicefrac{1}{g_{SU(3)_a}^2}=\nicefrac{1}{g_{SU(4)_h}^2}=\nicefrac{2}{3g_{USp(2)_b}^2}=\nicefrac{6}{19g_{U(1)_Y}^2}$,
 i.e. the $SU(3)_a$ and $SU(4)_h$ gauge couplings at tree-level at $M_{\text{string}}$ are slightly stronger than those of the $USp(2)_b$ and 
 $U(1)_Y$ factors.
 \item
The one-loop gauge threshold corrections from D6-branes at some vanishing angle displayed - for the annulus topology only - in table~\ref{Tab:Betas+1-Loop}
contain lattice sums, which have the asymptotic behaviour $\Lambda_{\tau,\sigma}(v) \stackrel{v \to \infty}{\longrightarrow}
\hat{c}_{\tau,\sigma} \frac{\pi v}{3}$ with $\hat{c}_{0,0}=-1$, $\hat{c}_{\tau,1}=1$ and $\hat{c}_{1,0}=-2$.
For highly unisotropic choices of two-torus volumes $v_i \sim v_jv_j$, the one-loop correction linear in the largest volume $v_i$ and with negative prefactor, i.e. 
relative displacement $\sigma^i=1$, can thus compensate the tree-level value $\nicefrac{1}{g_{a,\text{tree}}^2} \propto \sqrt{v_1v_2v_3}$.

In the MSSM example of section~\ref{Sss:MSSM-Spectrum}, the one-loop corrections of the QCD and hidden stack scale 
as $\delta^{\text{loop}}_{\cal A} ( \nicefrac{1}{g_{SU(3)_a/SU(4)_h}^2})  \stackrel{v \to \infty}{\longrightarrow} {\rm const.} \times \frac{6 v_1 \pm v_2}{6}$.
The degeneracy of the two gauge couplings is thus lifted by the one-loop gauge threshold correction. The $SU(4)_h$ coupling
becomes stronger, favouring the formation of a hidden sector gaugino condensate somewhere between $M_{\text{string}}$ und $M_{\text{weak}}$.
At this point, it is noteworthy that including contributions from M\"obius strip amplitudes 
 leads to additional corrections $\delta^{\text{loop}}_{\cal M} ( \nicefrac{1}{g_{SU(3)_a/SU(4)_h}^2}) \propto
 2 \Lambda_{0,0}(v_1) - \hat{\Lambda}_{1,1}(v_2) + \hat{\Lambda}_{1,1}(v_3)$ with unknown shape of the lattice sum $\hat{\Lambda}$ for tilted tori.
 Reliable field theoretical predictions are either possible in the regime $v_1 \gg v_2,v_3$ or if $v_2 \equiv v_3$.\cite{Honecker:2016-3}
\end{enumerate}
In summary, the string scale can - even in simple toroidal orbifold set-ups - be chosen considerably below the GUT scale, opening 
up new arenas for new physics scenarios. The decay constant of e.g. a closed string axion participating in the Green-Schwarz mechanism 
scales as $f_{\xi} \sim M_{\text{string}}$, and can such an axion provide an explanation for dark matter if $M_{\text{string}}$ lies in an intermediary 
range of about $10^{9-12}$ GeV, which is also of interest to SUSY breaking.
Going to the extreme case of $M_{\text{string}} ={\cal O}$(TeV) on the other hand would allow for $Z'$ bosons or dark photons 
that might become visible at the LHC or some future collider, see e.g.Refs.~\citen{Kiritsis:2003mc,Anchordoqui:2011eg,Cvetic:2011iq,Lebed:2011fw,Buckley:2011vc,Anchordoqui:2012wt,Berenstein:2014wva} for extended discussions.

\section{Conclusions and Outlook}\label{S:Conclusions}	

String theory as a framework for unifying General Relativity and QFT seemingly balances between ``anything goes'' and ``nothing goes''. 
Just as the Standard Model would be a very unlikely QFT if one were to take a statistical approach over all possible gauge groups and charged fermions, 
phenomenologically appealing embeddings of particle physics are very rare within the landscape of four-dimensional string vacua.
The related representation theory is very well understood and provides a valuabe guideline for possible BSM and dark sector components
compatible with some UV completion. Prominent examples are provided by open and closed string axions as well as massive gauge bosons.
The role of the (pseudo)scalars can range from acting as inflaton, solving the strong CP problem, forming dark matter components or dark radiation, see e.g. 
the recent article in Ref.~\citen{Ballesteros:2016euj} for an extended list of references.
Also massive gauge bosons can appear in various ways such as $Z'$ bosons or dark photons, which mediate between the visible and a hidden sector via kinetic mixing\cite{Abel:2008ai} with the hypercharge.

As argued here, the value of the string scale $M_{\text{string}}$ is strongly model-dependent within the framework of Type II orientifold compactifications.
Highly unisotropic compactifications in combination with one-loop corrections to the gauge couplings can in principle lower $M_{\text{string}}$
to an intermediary range of about $10^{9-12}$ GeV, or sporadically even to the TeV-range. In the latter case, an extreme amount of fine-tuning 
and stringy signatures at the LHC are to be expected.

While a classification in terms of gauge groups and matter representations seems to be within reach, deriving 
the four-dimensional effective action to a reliable level, where top-down scenarios of string-inspired field theoretical models 
of SUSY breaking, moduli stabilization or inflation can be tested, remains a major hurdle. 
As argued here, string phenomenology and string cosmology requires to carefully balance the beauty of geometrically engineering particle physics spectra
on generic compact backgrounds versus the accessibility of the related (e.g. gauge and Yukawa) couplings using SUGRA or, at special lampposts of 
the string landscape, CFT methods.
Concerted action in improving the knowledge of the low-energy effective action of string compactifications
will ultimately pave the way to the understanding of UV-consistent new physics phenomena in four spacetime dimensions.

\section*{Acknowledgments}

This work is partly supported by the {\it Cluster of Excellence PRISMA} DFG no. EXC 1098 and the DFG research grant HO 4166/2-2.

\bibliographystyle{ws-ijmpa}
\bibliography{refs_BriefReview-TypeII+BSM+DarkSector2016}

\end{document}